\documentclass[a4paper,twocolumn,aps,prb]{revtex4}
\usepackage{amsmath}
\usepackage{bbm}
\usepackage{amsfonts}
\usepackage{amssymb}
\usepackage{graphicx}
\usepackage{hyperref}
\usepackage{dsfont}
\usepackage{braket}
\usepackage{color}
\usepackage[T1]{fontenc}

\begin{document}

\author{Heng \surname{Wang}}
\author{Guido \surname{Burkard}}
\affiliation{Department of Physics, University of Konstanz, D-78457 Konstanz, Germany}
\date{\today}

\title{Creating arbitrary quantum vibrational states in a carbon nanotube}
\begin{abstract}
We theoretically study the creation of single- and multi-phonon Fock states and arbitrary superpositions of quantum phonon states in a nanomechanical carbon nanotube (CNT) resonator. In our model, a doubly clamped CNT resonator is initialized in the ground state and a single electron is trapped in a quantum dot which is formed by a electric gate potential and brought into the magnetic field of a micro-magnet. The preparation of arbitrary quantum phonon states is based on the coupling between the mechanical motion of the CNT and the electron spin which acts as a non-linearity. We assume that electrical driving pulses with different frequencies are applied on the system. The quantum information is transferred from the spin qubit to the mechanical motion by the spin-phonon coupling and the electron spin qubit can be reset by the single-electron spin resonance. We describe Wigner tomography which can be applied at the end to obtain the phase information of the prepared phonon states.  
\end{abstract}

\maketitle

\section{Introduction}

\begin{figure}[t]	
\begin{center} {\includegraphics[width=0.48\textwidth]{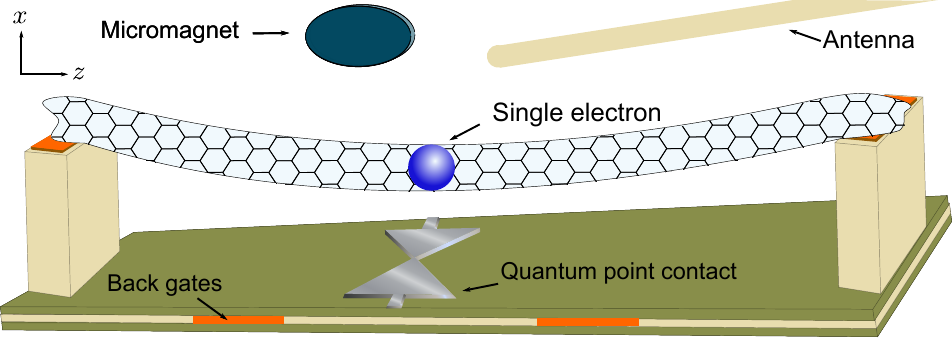}}  
\caption {Schematic view of a single electron being trapped in a quantum dot (QD) formed by gate voltages in a suspended carbon nanotube (CNT).   The resonance frequency of the CNT can be adjusted by the voltages on the back gate. An external ac electric field is applied on the CNT by the antenna. The micro-magnet is deposited in the vicinity of the CNT.  The single electron wavefuction can be electrostatically shifted by applying voltage on the back gates. Therefore the spin-splitting of the electron in the slanting magnetic field of the micro-magnet can be manipulated.\label{fig:setupcreat}} 
\end{center}
\end{figure}

The peculiar feature of quantum states is that they can be in a superposition of their basis states.
Preparation, manipulation and measurement of Fock states, which are quantum states with fixed numbers of quanta, and their superpositions are especially important for  quantum computation with trapped ions  \cite{Cirac1995}. 
The photon Fock states are widely used in quantum cryptography \cite{Bennett1984.,Beveratos2002}. 
The harmonic phonon states of a single trapped ion have been used as the control qubit with the hyperfine ground state as the target qubit in an experimental realization of two-qubit controlled-NOT quantum  gate \cite{Monroe1995}. The preparation of Fock states and their arbitrary superpositions in linear resonators has been proposed by transferring the quantum information of a nonlinear quantum system which can be controlled by a classical source \cite{Law1996} and has already been realized experimentally by coupling a trapped ion \cite{Meekhof1996} or a superconducting quantum circuit \cite{Hofheinz2008,Hofheinz2009}, to a resonator.  
Recently, single phonon states were proposed as qubit states in optomechanical schemes \cite{Stannigel2012}. 
Measurements of the quantum ground state and preparation of a single phonon state of a piezoelectric resonator coupled to a superconducting quantum bit have been achieved a few years ago  \cite{OConnell2010}. 
Heralded single-phonon preparation is obtained by detecting the photon of the photon-phonon pair generated by optomechanical parametric down-conversion \cite{Galland2014}.
The improvement of photon detection in the laboratory promises the precise single-photon counting  allowing for  single-phonon counting \cite{Cohen2015}. 
In nanomechanical or micromechanical systems, cooling the mechanical system \cite{Kippenberg2008,Marquardt2007} to the ground state and preparing nonclassical states are required to operate mechanical resonators in the quantum regime. Ground-state cooling of a mechanical system has been achieved with direct or active cooling in several laboratories \cite{Teufel2011,OConnell2010}.

Both electrical and mechanical properties of carbon nanotubes (CNTs) make them very interesting for quantum physics. 
Because of the additional valley degrees of freedom, semiconducting CNTs are promising candidates for valleytronics and valley-spin based technology  \cite{Rohling2014,Szechenyi2013,Palyi2010}. The curvature induced spin-orbit coupling in CNTs has been predicted to be significant \cite{Ando2000,Izumida2009,Jeong2009} and been observed in the laboratory \cite{Kuemmeth2008}. A magnetic field leads to the lifting of the four-fold spin and valley degeneracy \cite{Entin2001,Chico2004,Min2006,Huertas-Hernando2006,Chico2009,Kuemmeth2010}. 
On the other hand, suspended nanomechanical CNTs have high and widely tunable resonance frequencies and enormous quality factors \cite{Sazonova2004,Huettel2009,Chaste2011,Laird2012}, hence the vibrational modes of CNTs last long until they are totally damped out (Fig.~\ref{fig:setupcreat}).
The two lowest energy levels of anharmonic nanomechanical CNT oscillators have been proposed as the two states of one qubit in quantum information processing \cite{Rips2013}. The coupling of the electron spin and the mechanical motion of the CNT via the intrinsic spin-orbit coupling provides a non-linearity \cite{Rudner2010,P'alyi2012}. Many theoretical proposals for the read-out of the vibrational frequency of a suspended CNT \cite{Ohm2012a} and the electron spin states \cite{Struck2014}, for obtaining single- and two-qubit quantum gates \cite{WangandBurkard2014,Wang2015} and cooling a suspended CNT \cite{Stadler2014} are based on this spin-phonon coupling. Recently a theoretical work has proposed the ground-state cooling of a suspended carbon nanotube (CNT) resonator between a normal and superconducting lead by the interference of vibration-assisted Andreev reflections \cite{Stadler2015}.

We present theoretically how to prepare the Fock states and arbitrary quantum phonon states  based on the spin-phonon interaction in a suspended CNT. The basic working principle is similar to the one used previously for superconducting qubits \cite{Hofheinz2008,Hofheinz2009} and consists of the following steps. Two-electron spin states split by a magnetic field are defined as our qubit. The qubit flip, the qubit-phonon swap, and the phase operations are applied alternately to obtain an arbitrary quantum vibration state of the CNT. The qubit is flipped from the ground state $\ket{\downarrow }$  into the excited  state $\ket{\uparrow}$ by the electron spin resonance which is obtained in the presence of an external ac electric field matching the qubit frequency. The quantum dot is moved back and forth by the ac electric field hence the electron in the quantum dot experiences effectively a time-dependent magnetic field.  A qubit-phonon swap converts the energy from the excited qubit state to the resonator from the ground phonon state by the spin-phonon coupling. The qubit is brought into resonance with the phonon to have an effective spin-phonon coupling strength by electrostatically moving the quantum dot in the stray field of a micro-magnet \cite{Tokura2006,Pioro-Ladriere2007,Pioro-Ladriere2008}. 
A phase rotation of the spin can be applied to adjust the relative phase of the qubit. A sequence which alternates these three operations is applied until the desired quantum phonon state is obtained.

This paper is organized as follows. The quantum mechanical system and the effective Hamiltonian are introduced in Sec.~\ref{sec:modelFOCK}.  The respective time-evolution operators for the three necessary operations are presented in Sec.~\ref{sec:time-evolution}.  In Sec.~\ref{sec:superposition}, the steps to obtain Fock states and arbitrary quantum phonon states are explained. In Sec.~\ref{sec:wigner}, we  discuss the Wigner tomography for extracting the full information of the quantum phonon states. 
 
\section{Model}
\label{sec:modelFOCK}

We assume that a single electron is trapped within a quantum dot (QD) formed in a suspended CNT which lies between two supports (Fig.~\ref{fig:setupcreat}). The QD is controlled by voltages on the electrodes at the ends of the CNT. The resonance frequency of the CNT  $\omega_p$ can be adjusted by the back gates. The strength and the frequency of the electric driving field are denoted as $\lambda$ and $\omega$. In CNTs, there exists a curvature induced spin-orbit interaction which already splits the degeneracy of spin in each valley without any magnetic field.  With a magnetic field $B$ applied along the CNT, the four-fold energy degeneracy of the valley and the spin is completely lifted. The electrons in the $K$ and $K'$ valleys move in the directions of clockwise and anti-clockwise around the circumference of the CNT, respectively. Two spin states in the same $K'$  valley cross at the field $B^*=\Delta_{\rm SO}/2\mu_B$ where $\Delta_{\rm SO}$ is the spin-orbit interaction and $\mu_B$ is the spin magnetic moment. Since these two spin states are well separated in momentum space from the states in the other valley, we choose them as the qubit. The energy splitting of the qubit is $\hbar \omega_q=g_e\mu_B(B-B^*)$ where $B$ is the applied magnetic field,  $\omega_q$ denotes the qubit frequency and  $g_e$  is the electron $g$-factor \cite{Struck2014}. The micro-magnet, which produces a slanting magnetic field can be deposited near the CNT such that  the QD is located in the field. One can electrostatically move the QD and hence adjust the qubit frequency.  The frequency difference of the energy between the phonon and the qubit is denoted as $\Delta=\omega_p-\omega_q$. An external ac driving electric field is applied to the system for the spin flip  operation.

It is the spin-phonon interaction that converts the excitation of the qubit into quantum vibrational motion. The spin-phonon interaction applies with both of the deflection and the deformation phonon modes. In the following, we only consider a single polarization of the deflection mode of the CNT. It is possible to make  generalizations to other deflection modes and to the deformation modes. We assume that the resting CNT axis is along $z$ axis. The vibration of the CNT causes local changes in the direction of the CNT axis, and hence the tangent vector $\boldsymbol{t}(z)$ is dependent on the displacement $u(z)$ of the CNT. The interaction of the spin and the deflection phonon mode is induced by the spin-orbit interaction $H_{\rm SO}=\Delta_{\rm SO}\boldsymbol{\sigma}\cdot\boldsymbol{t}\simeq \Delta_{\rm SO}\sigma_{z}+\Delta_{\rm SO}( {\rm d}  u_{x}(z)/{ \rm d}  z) \sigma_x$. Here $\vec{\sigma}=(\sigma_x,\sigma_y,\sigma_z)$ is the vector of Pauli matrices, $u_x(z)$ is assumed as the displacement at point $z$ along the CNT in the $x$ direction and  $ u_x (z)\propto f(z)\frac{l_0}{\sqrt{2}}(a+a^\dagger)$ as a function of phonon creation and annihilation operators $a^\dagger$ and $a$, where $f(z)$ is the waveform of the QD and $l_0$ is the zero-point amplitude of the phonon mode \cite{P'alyi2012}. The spin-phonon interaction strength of the QD is $g=\Delta_{\rm SO}\braket{f'(z)}l_0/2\sqrt{2}$.
The Hamiltonian for this system is 
\begin{equation}
\begin{split}
&H =H_0 +H_{\rm d}+H_{\rm sp},\\
&  H_0 =   \frac{\hbar \omega
_{{\rm q} }}{2}\sigma_{z }+\hbar \omega_{\rm p}a^{\dag}a,\\
& H_{\rm d}=  2 \hbar \lambda(a+a^{\dag}) \cos(\omega t),\\
& H_{\rm sp}= \hbar g  (a+a^{\dag})(\sigma_{+ }+\sigma_{- }),
\end{split}
\label{eq:totalH}
\end{equation}
where $\sigma_{+}$ and $\sigma_{-}$ are qubit raising and lowering operators, respectively. Here, $H_0$ is the undisturbed Hamiltonian of the phonon mode and the electron spin qubit. $H_{\rm d}$ contains the external ac electric driving term where $\lambda$ is the driving strength and $\omega  $ is the driving frequency, and the third part $H_{\rm sp}$ denotes the spin-phonon coupling which is induced from the spin-orbit coupling.

\begin{figure}[t]	
\begin{center}
                       {\includegraphics[width=0.48\textwidth]{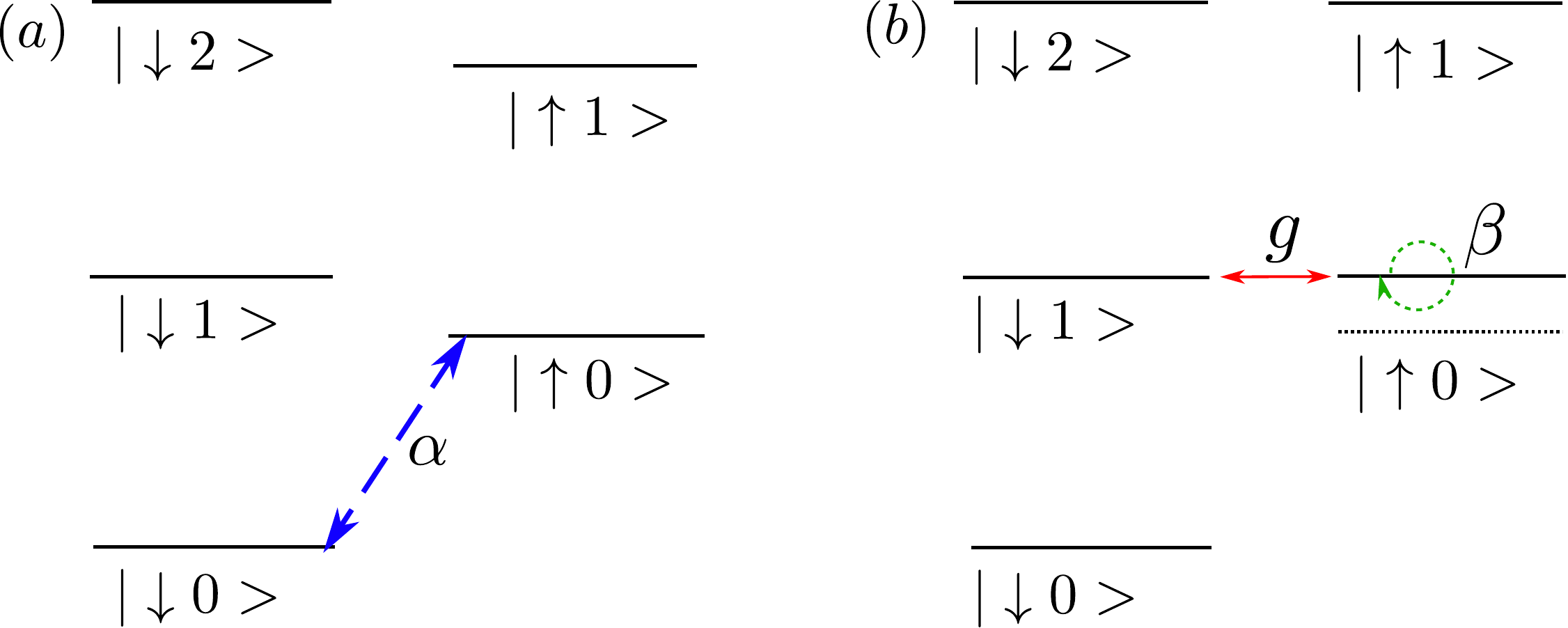}}                    
\caption {The energy-level diagram of the spin-phonon states. (a) The electron spin resonance between $\ket{\downarrow 0}$ and $\ket{\uparrow 0}$ (Blue dashed). The qubit is detuned from the phonon. The parameter  $\alpha$ is the effective strength  of the spin operator  $\sigma_x$ in  the effective Hamiltonian \eqref{eq:Heffective}. (b) The qubit is brought into resonance with the phonon in the slanting magnetic field of the micro-magnet by electrostatically moving the QD.  The spin-phonon interaction strength is $g$ (red solid). Phase operation  $\sigma_z$  with the strength $\beta$ is applied to adjust the relative phase of the state (green dotted). 
\label{fig:energylevel}}
\end{center}
\end{figure}

We assume that the detuning fulfills $\Delta \gg g, \lambda $. By applying a Schrieffer-Wolff transformation, an effective Hamiltonian from Eq.~\eqref{eq:totalH} is obtained in the interaction picture with respect to $H_0$ \cite{WangandBurkard2014}:
\begin{equation}
\begin{split}
&\tilde{H}'_I=  - \hbar\alpha\sigma_{x}+\hbar \beta_n\sigma_{z},
\end{split}
\label{eq:Heffective} 
\end{equation} 
where
\begin{equation}
\begin{split}
&\alpha=\frac{\lambda g \omega_p(\omega^2-2\omega^2_p+\omega_q^2) }{(\omega^2-\omega_p^2)(\omega_p^2-\omega_q^2)}, \\ 
&\beta_n=  \frac{1}{2}\omega_q  -\frac{1}{2}\omega_q\frac{2(2 n +1) g ^2 }{\omega_p^2-\omega_q^2} -\frac{1}{2}\omega.
\end{split}
\label{eq:Heffective1} 
\end{equation}
The eigenstates after the Schrieffer-Wolff transformation are slightly different from the original states because the higher order terms in the approximation are omitted. In the effective Hamiltonian, the term $\sigma_x$ ($\sigma_z$) denotes a rotation of the spin about the $x$ ($z$) axis of the qubit. We can obtain one of these two spin rotations separately by setting the   coefficient of the other rotation to zero. For example, to obtain a rotation about the axis $x$, we set $\beta_n$ to zero as shown in Fig.~\ref{fig:energylevel}.  The rotations about the $x$ axis can be  used for obtaining electron spin resonance (ESR) and  flipping the qubit states in the preparation of the arbitrary quantum phonon states. The rotations about the $z$ axis can be used as a phase operation. Here, $n$ denotes the phonon number.

The spin-phonon interaction is used to exchange the information between the qubit and the phonon where the driving field is off. The Hamiltonian without the external driving $H_{\rm d}$ in the rotating wave approximation in the interaction picture with respect to $H_0$ is 
\begin{equation}
\begin{split}
& H''_I= -\hbar\Delta\sigma_{z }+\hbar g (a\sigma_{+ } +a^{\dag}\sigma_{- }).
\label{eq:Hamiltonianswap}
\end{split}
\end{equation}
If a large detuning $\Delta$ is present, the effective coupling between the spin and the phonon is too small to convert the energy from the qubit to the resonator. To obtain a perfect swap of the qubit and the phonon, one can tune the frequency of the qubit to be in resonance with the phonon as shown in Fig.~\ref{fig:energylevel}. Together with the slanting magnetic field of a micro-magnet, electrostatically tuning the electron wave function of the QD serves this purpose \cite{Tokura2006,Pioro-Ladriere2007,Pioro-Ladriere2008}.

\section{Time-evolution operators}
\label{sec:time-evolution}

Since we have the effective Hamiltonian for the qubit flip and the phase operations in Eq.~\eqref{eq:Heffective},   and for the spin-phonon swap in Eq.~\eqref{eq:Hamiltonianswap}, we can derive their time-evolution operators with the aim of calculating the sequence of pulses for obtaining arbitrary quantum phonon states.

The interaction Hamiltonian in Eq.~\eqref{eq:Heffective} can be written as $\tilde{H}'_{I}=\boldsymbol{b}\cdot\boldsymbol{\sigma}$. 
The time-evolution operator of  the ESR  for the qubit flip operation, which is obtained by $e^{-i \boldsymbol{b} \cdot \boldsymbol {\sigma} t}=\cos(|\boldsymbol{b}|t)\mathbbm{1}-i \sin(|\boldsymbol{b}|t) (\hat{\boldsymbol{b}} \cdot \boldsymbol{\sigma})$,  with the Hamiltonian in Eq.~\eqref{eq:Heffective1} in the basis $\{\ket{g n},\ket{e n}\}$ with the phonon number $n$, is found to be
\begin{equation} 
\begin{split}
&R_n=e^{-i\tilde{H}'_{I}t/\hbar} \\
=&\left( 
\begin{array}{cc}
\cos ( \vartheta t )+i \frac{\beta_n}{\vartheta} \sin( \vartheta t)& i \frac{\alpha}{\vartheta} \sin(  \vartheta t)  \\
i \frac{\alpha}{\vartheta} \sin(  \vartheta t) & \cos ( \vartheta t )-i \frac{\beta_n}{\vartheta} \sin( \vartheta t)
\end{array} \right),
\end{split}
\label{eq:TEOtotal}
\end{equation}
where $\vartheta= \sqrt{\alpha^2 +\beta_n^2}$. 

We can obtain the time-evolution operator of the phase gate with $\alpha=0$ and $\omega=0$ in Eq.~\eqref{eq:Heffective1} and the electron spin rotates about the $z$ axis in the magnetic field. Hence we obtain the phase operation, 
\begin{equation} 
\begin{split}
P_n=e^{-i\tilde{H}'_{I}t/\hbar}  
=\left( 
\begin{array}{cc}
e^{i \beta_n t }& 0 \\
0 & e^{-i \beta_n t}
\end{array} \right),
\end{split}
\end{equation} 
where  the coefficients $\beta_n $ are different for the phase operators with different phonon numbers.

The time-evolution operator for the qubit-phonon swap with the Hamiltonian in the Eq.~\eqref{eq:Hamiltonianswap} in the basis of $\{ \ket{n \uparrow},\ket{n+1 \downarrow}  \}$ with $n=0,1\ldots$ is 
\begin{equation} 
\begin{split}
&U_n =e^{-i\tilde{H}''_{I}t/\hbar}  \\ & =\left( 
\begin{array}{cc}
\cos (g\eta_n t)+i \Delta \frac{\sin(g\eta_n t)}{\eta_n} & -i \sqrt{n+1} \frac{\sin(g \eta_n t)}{\eta_n} \\
-i \sqrt{n+1} \frac{\sin(g\eta_n t)}{\eta_n}   &  \cos (g\eta_n t)+i\Delta \frac{\sin(g\eta_n t)}{\eta_n}
\end{array} \right),
\end{split}
\end{equation}
where $ \eta_n= \sqrt{n+1+ \Delta^2/g^2} $. 
The swap between the qubit and the phonon can be achieved best when they are on resonance. For the resonant case $\Delta=0$, we have a simple time-evolution operator of the qubit-phonon swap 
\begin{equation} 
\begin{split}
U_n=e^{-i\tilde{H}''_{I}t/\hbar}  =\left( 
\begin{array}{cc}
\cos (g t\sqrt{n+1} ) & -i    \sin(gt \sqrt{n+1} )  \\
-i  \sin(g t\sqrt{n+1})   &  \cos (gt\sqrt{n+1} )
\end{array} \right).
\end{split}
\end{equation}

It is worth pointing out that the qubit flip and the qubit-phonon swap both depend on the phonon numbers. States with different phonon numbers have different coefficients hence require different times for the same  swap or flip operations. For example, the swap operations of $\ket{\downarrow 1}\rightarrow\ket{\uparrow0}$ and $\ket{\downarrow 2}\rightarrow\ket{\uparrow1}$ require different times because the phonon numbers are different.  This leads to dephasing in the electronic sector in the  state  preparation  protocol. 
The dephasing can be canceled in the process of preparation by applying uncompleted swap and flip operations together with phase operations.

\section{Arbitrary quantum phonon states}
\label{sec:superposition}

\begin{figure}[t]	
\begin{center} {\includegraphics[width=0.48\textwidth]{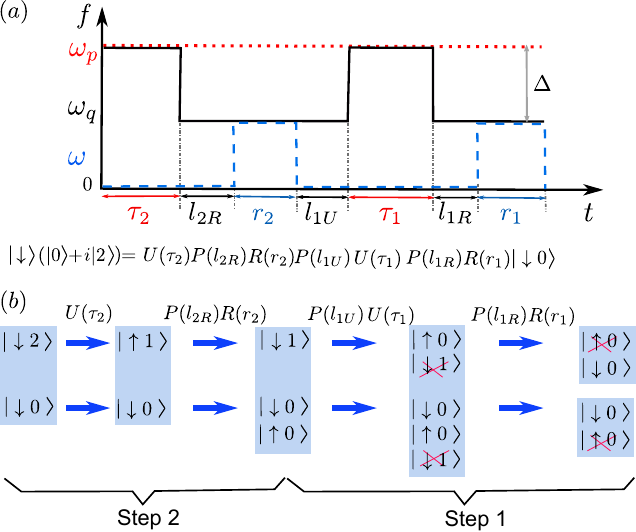}}              
\caption {(a) Sequence of operations for obtaining arbitrary quantum phonon states. An external ac electrical field with the frequency $\omega$ (blue dashed) is applied for obtaining the qubit flip in the time intervals $r_1$ and $r_2$, and $\omega_p$ is the frequency of the phonon mode (red dotted) in the CNT. The large detuning $\Delta$ is required for the qubit flip  operation  $R$. For the qubit-phonon swap $U(\tau)$, the qubit with the frequency $\omega_q$ (black solid) is brought into resonance with the phonon that $\omega_p=\omega_q$ by moving the QD in the slanting magnetic field of the nearby micro-magnet. The swap operation's time intervals are $\tau_1$ and $\tau_2$. The phase operations $P(l_{1U})$, $P(l_{1R})$ and $P(l_{2R})$ adjust the relative phase of the state. (b) Diagram for calculating the operation sequence in a backwards direction for obtaining the superposition of Fock states $\ket{\downarrow}(\ket{0}+i\ket{2})$ from the ground state $\ket{\downarrow 0}$. The operation  $U(\tau_2)$ is applied to fully transfer the state $\ket{\downarrow 2}$ to the  state $\ket{\uparrow 1}$ and the flip operation $R(r_2)$ is applied to fully transfer the state $\ket{\uparrow 1}$ to $\ket{\downarrow 1}$ in step No.~2. In step No.~1, phase operations are applied to adjust the relative phases to cancel some states (red crosses), e.g. the states $\ket{\downarrow 1}$ and $\ket{\uparrow 0}$  in the following qubit flip or qubit-phonon swap.
\label{fig:pulses}}
\end{center}
\end{figure}

Now we explain the operation sequence of obtaining arbitrary phonon Fock states and  superpositions of phonon Fock states. 
To obtain a phonon Fock state  $\ket{\psi_n} =\ket{\downarrow n}$ from  the ground state $\ket{\downarrow 0}$, a sequence of operations  with   the qubit-phonon swaps and qubit flips for $n$ steps is applied as $$\ket{\psi_n} = U(\tau_n)R(r_n)...U(\tau_1)R(r_1)\ket{\downarrow 0}$$ where $\tau=\frac{\pi/2}{g \sqrt{n}}$ and $r=\frac{\pi/2}{\alpha}$ are the timescales of operations. Here we assume $\beta_n=0$ in qubit flip operators $R$.
Each step contains a qubit-phonon swap and a qubit flip and the highest phonon number increases by $1$ after each step.
By applying the ac electrical field in the presence of the large detuning, the qubit flips from $\ket{\downarrow n}$ to $\ket{\uparrow n}$ completely in each qubit flip operation. The qubit-phonon swap transfers the energy completely from the excited spin state to the resonator, i.e. from $\ket{\uparrow n}$ to  $\ket{\downarrow n+1}$.

To obtain the arbitrary phonon state $\ket{\psi} =\sum_n c_n \ket{\downarrow n}$, a sequence of operations with $n$ steps is applied on the initial state  $\ket{\downarrow 0}$ as 
\begin{align}
\ket{\psi} = &  U( \tau_n) P(l_{nR})R(r_n) \ldots\\
&\ldots P(l_{1U})U(\tau_1))P(l_{1R})R(r_1)\ket{\downarrow 0}.
\end{align}
A sequence of one qubit flip $R$, one qubit-phonon swap operation $U$, and two phase rotation operations  $P$  is applied in each step,   except only one phase rotation is applied in the last ($n$-th) step. The sequence is calculated backwards from the target state to the ground state $\ket{\downarrow 0}$. Each step decreases the highest phonon number by $1$.  We can apply one step of the sequence of operations for obtaining the state $ \ket{\downarrow}(\ket{0}+ \ket{1})$ as 
\begin{align}
\ket{\downarrow}&(\ket{0}+ \ket{1}) =
U(\tau_1)P(l_{1R})R(r_1)\ket{\downarrow 0}. 
\end{align}
Here we apply a complete swap $U(\tau_1)$ and an uncompleted qubit flip $R(r_1)$. The phase operator $P(l_{1R})$ is applied to regulate the relative phase of the state. The state $\ket{\downarrow}(\ket{0}+ \ket{1})$ will transfer into $(\ket{\uparrow}+\ket{\downarrow})\ket{0}$ under the complete swap operation. The spins $(\ket{\uparrow}+\ket{\downarrow})\ket{0}$ are flipped in the uncompleted qubit flip operation $R(r_1)$ so that only $\ket{\downarrow 0}$ is left, while a complete qubit flip would lead to an unwanted state $\ket{\uparrow 0}$ which causes dephasing.  We assume the parameter $\beta_n=\beta=0$ in the qubit flip  operator $R$.  The analytical expressions of the operation times are obtained as $r_1=\frac{-\frac{3\pi}{4}+2\pi C_1 }{\alpha}$, $l_1=-\frac{2 i (\omega_p^2-\omega_q^2)(i \pi+2i \pi C_2)}{\omega_q(2g^2-\omega_p^2+\omega_q^2)}$ and $\tau_1=\frac{\pi/2+2\pi C_3}{g}$, where $C_{ i=1,2,3}$ are non-negative integers.


To explain how to apply the sequence of operations, we consider an example of obtaining the state $ \ket{\downarrow}(\ket{0}+i \ket{2})$. As shown in Fig.~\ref{fig:pulses}, the sequence is calculated in the time reversed order from  $ \ket{\downarrow}(\ket{0}+i \ket{2})$ to $\ket{\downarrow 0}$ and we obtain 
\begin{align}
\ket{\downarrow}&(\ket{0}+i \ket{2}) = \\
&U( \tau_2) P(l_{2R}) R(r_2) P(l_{1U})U(\tau_1))P(l_{1R})R(r_1)\ket{\downarrow 0}. 
\end{align}
 Fig.~\ref{fig:pulses} (a) shows the frequencies of the phonon $\omega_p$, the qubit $\omega_q$ and the driving $\omega$ as a function of the time. We can see that the qubit frequency is brought  into resonance with the phonon frequency during the qubit-phonon swap. In the qubit flip  operation, the driving is applied and a large detuning $\Delta$ of the qubit frequency and the phonon frequency is required.  Completed operations of the qubit-phonon swap  and the qubit flip are applied in order to decrease the highest phonon number by one in step No.~$2$.  In qubit flip operation $R(r_2)$, the state  $\ket{\uparrow 1}$ flips completely to $\ket{\downarrow 1}$.  However, due to the spin-phonon coupling strength and the spin flip strength both depend on phonon numbers,  dephasing of the states with lower phonon numbers appear in the  process. Here since the qubit flip  depends on phonon numbers, the state $\ket{\downarrow 0}$ could not fully flip to the state $\ket{\uparrow 0}$ in time $r_2$ therefore causes the dephasing in the electronic sector. The remaining state $\ket{\uparrow 0}$ could be swapped to the state $\ket{\downarrow 1}$ in the next qubit-phonon swap  operation, which would be with the highest phonon number, therefore we want to cancel $\ket{\uparrow 0}$ to avoid this. To cancel these dephasing we apply phase operations to adjust the relative phase of the state and perform uncompleted  qubit flips and qubit-phonon swaps. The phase rotations are necessary when the next swap or flip  operations are not applied completely. When a spin up state and the Fock state with the highest phonon number need to be canceled, we apply the phase operation and an uncompleted qubit flip. To cancel a spin down state with the highest Fock state, we apply a phase operation and a qubit-phonon swap  operation. Therefore in step No.~$1$ the  phase operator $P(l_{1u})$ is applied to adjust the relative phase of the state and the qubit-phonon swap $U(\tau_1)$ is applied partially   to  cancel  $\ket{\downarrow 1}$. Hence we have only $\ket{\uparrow 0}$ and the leftover state $\ket{\downarrow0}$ and the rest of the  operation is similar with the relevant part in the preparation of $\ket{\downarrow}(\ket{0}+\ket{1})$. 
 For obtaining other superpositions of Fock states with larger highest phonon numbers $n$ or  with more than two Fock states, one repeats the second step $n$ times.

\section{Wigner tomography}
\label{sec:wigner}

 \begin{figure}[b]
 \begin{center}
               {\includegraphics[width=0.48\textwidth]{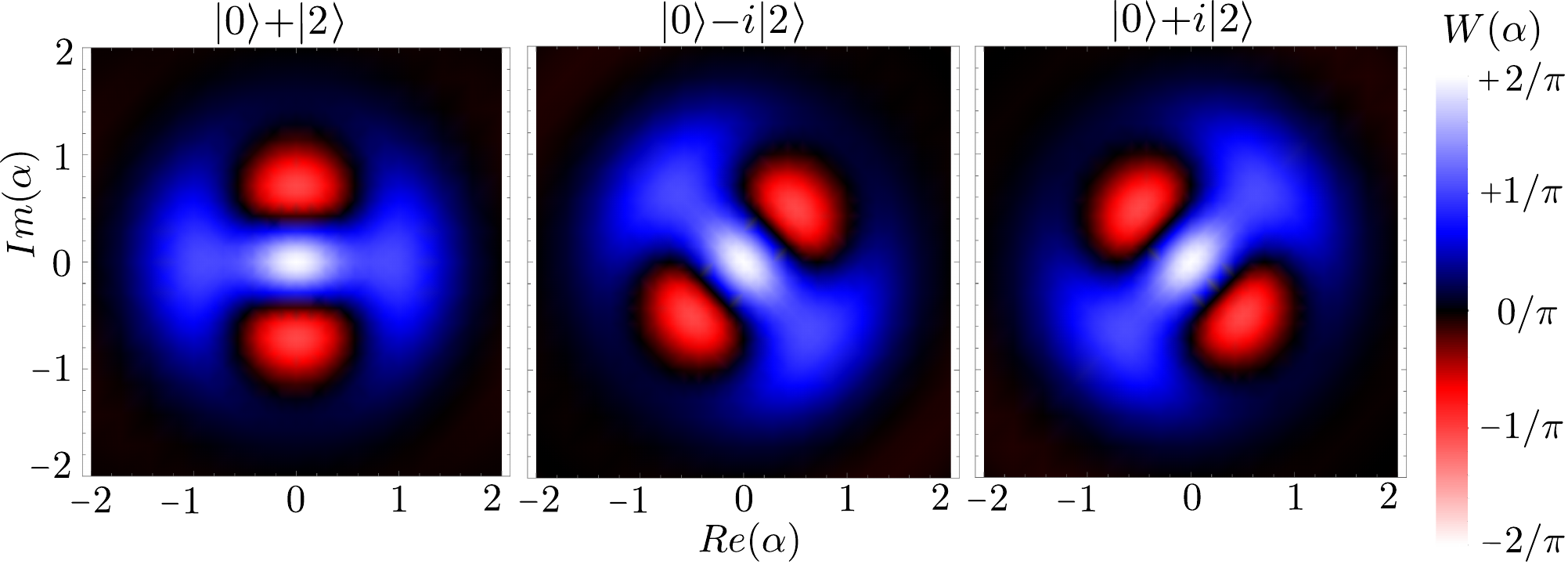}}         	     
\caption { The Wigner tomography of the quantum phonon states  $\ket{0}+\ket{2}$, $\ket{0}-i\ket{2}$ and $\ket{0}+i\ket{2}$. The change of the relative phase of a two-state superposition of Fock states rotates the Wigner function $W(\alpha)$. 
\label{fig:wigner0and2phase} }
\end{center}
\end{figure}

\begin{figure}[t]	\begin{center}
                    	  {\includegraphics[width=0.48\textwidth]{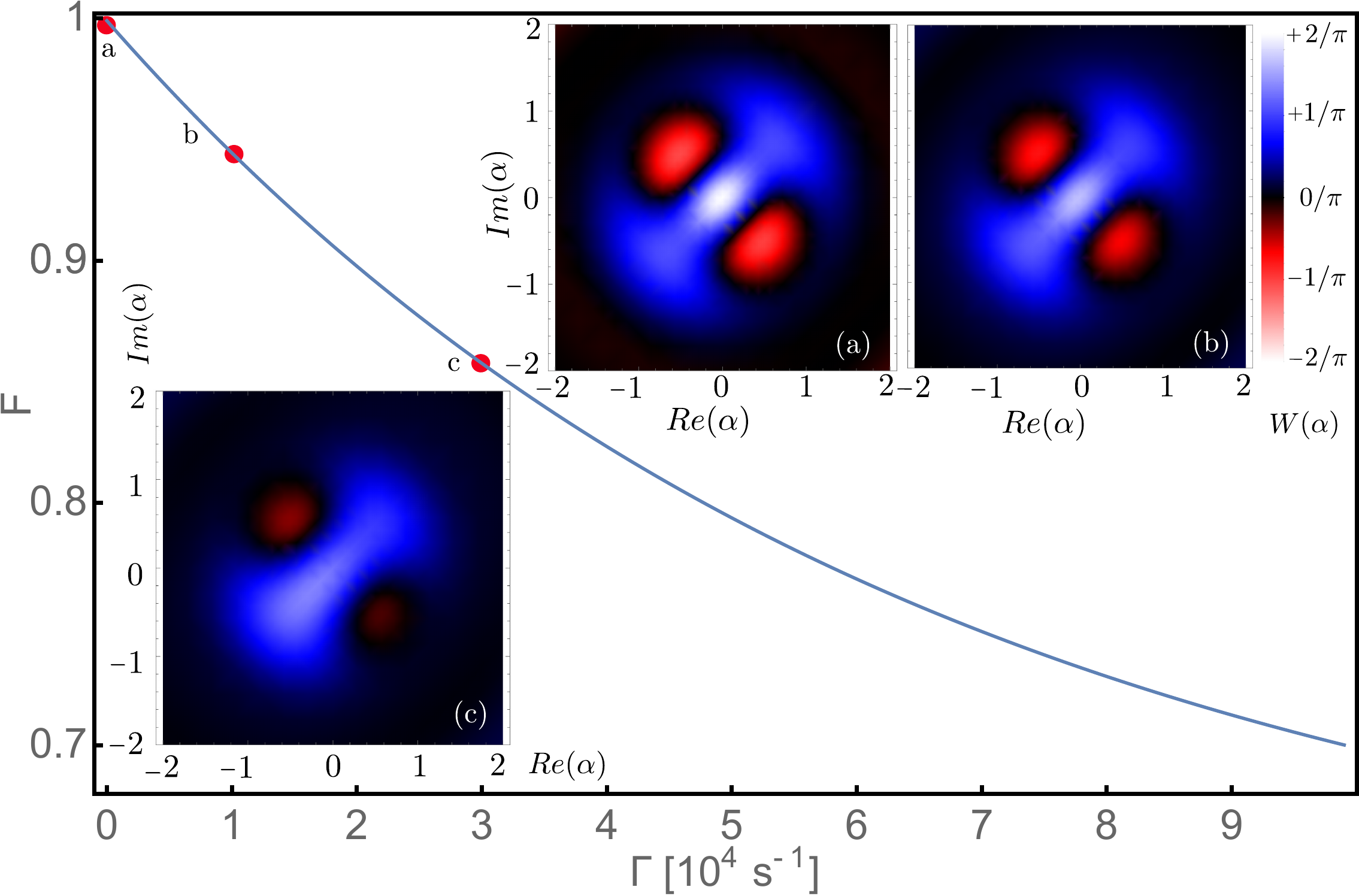}}               	               
\caption { The fidelity $F$ of obtaining  $\ket{\psi}=\ket{0}+i\ket{2}$  at temperature   $T=10 \  {\rm mK}$ as a function of the damping rate $\Gamma$.  The fidelity $F=\sqrt{\bra{\psi} \rho \ket{\psi}}$ shows how close the obtained state $\rho$ is to the target state $\ket{\psi}$. The Wigner tomography of the obtained states at $T=10 \  {\rm mK}$ with (a) damping $\Gamma=0$, (b)  $\Gamma=10^4  \ {\rm s^{-1}}$ and (c) $\Gamma=3\times 10^4  \  {\rm s^{-1}}$. The fidelity for the state obtained at (a) is $F =0.999$, for (b) is $F=0.945$ and for (c) is $F=0.859$. The other parameters are $\lambda/2\pi = 0.8\ {\rm MHz}$, $\Delta=100 \ {\rm MHz}$, $\omega_p/2\pi = 1.5\ {\rm GHz}$,  and $g/2\pi =0.56 \ {\rm MHz}$.   
\label{fig:fidelity} }
\end{center}  
\end{figure}


One can use  Wigner tomography \cite{Englert1989,Moya-Cessa1993,Banaszek1996,Haroche2006, Eichler2012} to obtain the relative phase of the quantum phonon states which has been used for quantum photon states \cite{Hofheinz2008,Hofheinz2009}. 
Wigner tomography is based on representing the Wigner function as a quasiprobability distribution on the complex phase space. The Wigner function can be written as the expectation value of the operator $D^\dagger(-\alpha)\varPi D(-\alpha)$ \cite{Hofheinz2009},
\begin{equation} 
 W(\alpha)=\frac{2}{\pi}\braket{\psi|D^{\dagger}(-\alpha)\varPi D(-\alpha)|\psi}.
 \label{eq:Wigner1}
\end{equation}
To obtain $D(-\alpha)$, the resonator is driven with an ac electric field pulse as $-\alpha=(1/2)\int \lambda(t) {\rm d}t$, where $\alpha$ is the phase space amplitude of the resonator and $D$ is the displacement operator $D(-\alpha)=D^\dagger(\alpha)={\rm exp}(\alpha^*a-\alpha a^\dagger)$. For the parity operator $\varPi$, Fock states have eigenvalues $1$ and $-1$ for even and odd phonon numbers, respectively. For mixed states, Eq.~\eqref{eq:Wigner1} can be written as a trace  
\begin{equation} 
\begin{split}
  W(\alpha)=&\frac{2}{\pi}{\rm Tr} ( D(-\alpha) \rho D(\alpha) \varPi)\\
  =&\frac{2}{\pi}\sum_n (-1)^n \rho'_{nn}(-\alpha),
 \label{eq:Wigner2}
\end{split}
\end{equation}
where $\rho$ is the density matrix $\rho=\sum_i P_i \ket{\psi_i}\bra{\psi_i}$ of the resonator before being displaced  \cite{Hofheinz2009}.  For the displaced resonator, the density matrix is $\rho'=D(-\alpha) \rho D(\alpha)$. To calculate the Wigner function, we need to obtain the phonon numbers $\rho_{nn}$ from the probability $P_n$ \cite{Lougovski2003}.
In principle, the phonon number $n$ can be measured directly with charge detector \cite{Struck2014}, but for small numbers of phonons in the CNT the accuracy is limited.
 After the displacement pulse, one brings the qubit on resonance with the resonator for a variable time and then  performs the read-out of the qubit.  The qubit can be read out by the mechanical response of the  resonator to the pulsed external driving \cite{Struck2014}. States with different spin states react to the external driving differently such that the excited spin states can be driven to other states with larger phonon numbers. Therefore one can tell apart the spin states by measuring the amplitude of the resonator via a charge detector. From the probability $P_u(t)$ of finding the qubit in state $\ket{\uparrow}$, we can obtain the measured probability for being into Fock state $\ket{n}$ as $P_n=|c_n|^2$ \cite{Hofheinz2008,Hofheinz2009}. The Wigner function rotates with the changes of the relative phase of a two-state superposition of Fock states as shown in Fig.~\ref{fig:wigner0and2phase}. For superpositions of more than two Fock states, the shapes of Wigner functions change.

 We can simulate a full set of measurements with probability $P_n$ for  having Fock state $\ket{n}$ via the density matrix $\rho$ from the set of the linear equations  \cite{Hofheinz2009}
 \begin{equation}
  \rho_{nn}'(\alpha)=\bra{n} D(-\alpha)\rho D(\alpha)\ket{n}=\sum_{j,i}M_{nji}\rho_{ji},
 \end{equation}
where the matrix $M$ has the form 
\begin{equation}
 M_{nji}=\bra{j}D(\alpha)\ket{n}^*\bra{i}D(\alpha)\ket{n}.
\end{equation}
The displacement operator can be expanded in the basis of Fock states as 
\begin{equation}
 \bra{u}D(\alpha)\ket{v}=e^{-|\alpha|^2/2} \sqrt{u!v!} \sum_{k=0}^{{\rm min}\{ u,v\}} \frac{\alpha^{u-k} (-\alpha ^*)^{(v-k)}}{k!(u-k)!(v-k)!}.
\end{equation}
Therefore we can obtain the Wigner function from the density matrix $\rho$, which is used in the following simulation.

\section{nonunitrary evolution}

We use a master equation for the  non-unitary evolution taking the damping of the CNT and the thermal bath into account. The spontaneous qubit relaxation rate is neglected due to the small density of other phonon modes which have similar frequencies in the CNT and in the surroundings such as the substrate and the supports. 
The master equation for the density matrix $\rho$ is of the form
 \begin{equation}
	\begin{split}
	\dot{\rho}=& -\frac{i}{\hbar}[H,\rho] +(n_{B}+1)\Gamma \left(a\rho a^{\dagger} -\frac{1}{2}\{a^{\dagger} a,\rho\}\right) \\
	& +n_{B}\Gamma\left(a^{\dagger}\rho a-\frac{1}{2}\{a a^{\dagger},\rho\}\right),
        \end{split}
\end{equation}
where $n_{B}=1/(e^{\hbar \omega_{\rm p}/k_{\rm B}T}-1)$ is the Bose-Einstein occupation factor and  $\Gamma \ll g$ is the damping rate of the CNT. CNTs with high factor $Q=\omega_q/\Gamma\approx150000$ have been found in laboratories \cite{Huettel2010,Cirio2012}. We take the following parameters: $\Gamma=10^4 \ {\rm s^{-1}}$, $Q=950 000$, and $\omega_p/2\pi=1.5 \ {\rm GHz}$.  The phonons follow the Bose-Einstein statistics in the thermal equilibrium, $ \rho=\frac{1}{Z}\sum_{n=0}^{ \infty} e^{-n\hbar \omega_{\rm p}/k_{\rm B}T}\ket{n}\bra{n}\otimes\ket{\psi }\bra{\psi }$  where $Z=\sum_{n=0}^{ \infty} e^{-n\hbar \omega_{\rm p}/k_{\rm B}T}$ is the partition function.
We obtain the total phonon state  by the partial trace over the spins $\rho_{\rm ph}={\rm Tr}_{\rm s}\rho$. We have simulated a procedure to produce the state $\ket{\psi}=\ket{\downarrow}(\ket{0}+i\ket{2})$ at finite temperature $T=10 \ {\rm mK}$. The Fig.~\ref{fig:fidelity} shows how the fidelity $F=\sqrt{\bra{\psi} \rho \ket{\psi}}$ decreases with the damping rate at finite temperature.
  The fidelity  for the state obtained at $\Gamma=0 $ is $F_{1}=0.999$, and $F_{2}=0.945$ with the damping rate $\Gamma = 10^{4} \ {\rm s^{-4}}$, and the fidelity is found to be $F_3=0.859$ with the damping rate $\Gamma=3\times10^4 \ {\rm s^{-1}}$. 

\section{Conclusion}

In conclusion,  single Fock states and arbitrary superpositions of the Fock states can be obtained by  sequences of qubit-phonon swaps, qubit flips, and phase operations. The exchange of the spin and the phonon is obtained by the spin-phonon interaction, which is based on the coupling of the phonon and the spin due to the intrinsic spin-orbit interaction. To obtain a large spin-phonon coupling strength it requires the resonance of  the spin and the phonon.  The mechanically induced ESR, which is obtained by applying a external ac electric field, is used to flip the qubit in the presence of a large detuning of the qubit and the phonon. The frequency of the qubit can be adjusted by electrostatically moving the electron wave function in the CNT in the slanting magnetic field of a nearby micro-magnet.  A phase operation  is applied to change  the relative phase of the state to cancel unwanted Fock states in  the next qubit-phonon swap or  the next qubit flip. Wigner tomography can be used to obtain the phase and the amplitude information of the states. Non-unitary evolution of the system is simulated with the master equation.  Our proposal introduces a way of electrically creating arbitrary quantum phonon states by interacting the CNT resonator with the electron spin in CNT.   
The formation of maximally entangled quantum phonon states between two modes of a mechanical resonator can be further studied by transferring the information from two coupled electron spins in two quantum dots to the resonator or coupling one spin to two different modes.   

\bibliographystyle{apsrev}
\bibliography{creating}

\end{document}